\documentstyle[aps]{revtex}

\begin{document}
\draft
\preprint{}
 
\title{Vortex Motion and Vortex Friction Coefficient
in Triangular Josephson Junction Arrays}

\author{Wenbin Yu and D. Stroud\\
{\it Department of Physics, The Ohio State University, Columbus, OH 43210}}

\date{\today}
\maketitle

\begin{abstract}
 
We study the dynamical response of triangular Josephson junction
arrays, modelled as a network of
resistively- and capacitively-shunted junctions (RCSJ's).  A flux flow 
regime is found to extend between a lower vortex-depinning current and a 
higher critical current, in agreement with previous calculations for
square arrays.  The upper current corresponds either to row-switching 
events accompanied by steplike jumps in the array resistance, or to a 
depinning of the entire array.
In the flux flow regime, the dynamical response to the bias current is 
roughly Ohmic, and the time-dependent voltage  
can be well understood in terms of vortex degrees of freedom.
The vortex friction coefficient $\eta$ depends 
strongly on the
McCumber-Stewart parameter $\beta$, and at large $\beta$ is approximately 
independent of the shunt resistance $R$.  To account for this, we 
generalize a model of Geigenm\"{u}ller {\it et al} to treat energy
loss from moving vortices  
to the phase analog of optical spin waves in a triangular lattice.
The value of $\eta$ 
at all values of $\beta$ agrees quite well with this model in the 
low-density limit.  The vortex depinning current is estimated as $0.042I_c$, 
independent of the direction of applied current, 
in agreement with static calculations by
Lobb {\it et al}.  A simple argument suggests that quantum effects in vortex 
motion may become important when the flux flow resistivity is of order
$h/(2e)^2$ per unit frustration.
\end{abstract}


\narrowtext
\twocolumn
 
\maketitle
 
\newpage
 
\section{Introduction}

The behavior of vortices in Josephson junction arrays (JJA's) has attracted 
much recent attention.$^{1-14}$  Such
vortices are coherent arrangements of
phases of the superconducting order parameter, which may move through 
the array like particles, in response to forces generated by 
external currents.  They can be generated by a magnetic field, or  
excited thermally.   At low velocities, it has been proposed that the 
motion of a vortex in a square array can be described by a Josephson-like 
equation of the form$^3$
\begin{equation}
   \frac{d^2}{dt^2}\left(2\pi\frac{x}{a}\right) + \frac{1}{RC}\frac{d}{dt}
   \left(2\pi\frac{x}{a}\right) + \frac{4e}{\hbar C}
\left[I_d\sin \left(2\pi\frac{x}{a}\right)-I\right]=0.
\end{equation}
Here $x$ is the vortex position along a line through the plaquette centers
and perpendicular to the external current $I$,
$a$ is the lattice constant, $R$ and $C$ are the shunt resistance and shunt
capacitance of each junction, and $I_d$ is the vortex depinning
current, which is linearly related to the junction critical current
$I_c$.   This equation may describe the behavior of a vortex in a 
square array at low velocities.

Because of the mass term in this equation, a vortex might be expected 
to move ballistically under appropriate circumstances -- that is, 
a vortex, once set into motion, would remain in motion even if the driving 
current is turned off.  There are several experimental reports 
of such motion$^{4,6}$.   However, numerical calculations and analytical 
studies, based on 
{\em classical} equations of motion, have not yet produced 
such ballistic motion$^{7,9,10,14} $.  The numerical studies have been 
carried out either in square arrays$^{7,9,14}$, or using a 
simplified representation of the Josephson interaction, in triangular 
arrays$^{7}$.  Instead of ballistic motion, in these simulations, 
it is generally found that the junctions in the wake of the vortex oscillate,
causing the vortex to lose its energy to the array.  In an analytical study 
based on a continuum model$^{10}$, an equation of motion for individual 
vortices is derived for both square and triangular arrays, starting from an 
effective action for the array.  This paper concludes that a narrow window 
of vortex velocity exists in a triangular array, for which ballistic motion 
may be possible.  However, it seems unlikely that this window is the regime 
which is probed experimentally$^{4,6}$

In this paper, we present dynamical calculations for 
triangular Josephson junction arrays.  The calculations are carried out
within a classical model of resistively- and capacitively-shunted junctions 
(RCSJ's).  Our principal aim is to study the motion of single vortex in the 
presence of an applied d.\ c.\ current, to see if
ballistic vortex motion is possible. 
Since the depth of the vortex potential in a triangular array is believed to
be smaller than that of a square array$^{16}$, such motion would seem more 
likely, at first glance, than in square arrays.   However, we find no such 
ballistic motion under any circumstances investigated.  Instead,
the motion of the vortex, when it is a coherent excitation of the
array, generally falls into a ``flux flow'' regime, where
the vortex moves with approximately Ohmic resistivity, described by a 
characteristic vortex viscosity. 

As in previous simulations in square arrays$^{7, 9, 13, 14}$, and
in similar studies in triangular arrays based on a simplified Josephson
coupling$^{7}$, we find that the flux flow region terminates at 
sufficiently high current in either of two ways.  
One possibility is for the entire lattice of Josephson 
junctions to be ``depinned'' causing the voltage to increase sharply.
A second decay mode, which predominates in sufficiently underdamped arrays,
is for the vortex to excite row switching events, which produce steplike 
increases of the array resistance as an entire row of junctions switches 
from superconducting to normal.  Both such decay modes have been 
in a variety of experiments$^{5,8,17,18}$.
 
Perhaps the most striking result of our simulations, also reported 
previously in square arrays$^{9,14}$, is the persistence of 
the quasi-ohmic flux-flow regime even in high-resistance arrays where, 
according to simple models, ballistic motion should be possible.  To 
account for this, we generalize a model of 
Geigenm\"{u}ller {\it et al}$^{9}$ to obtain a simple, nearly analytic model 
from which an effective vortex friction coefficient can be computed at any 
value of the vortex velocity and junction McCumber-Stewart coefficient$^{19}$
\begin{equation}
   \beta =2eR^2I_cC/\hbar.
\end{equation}
$\beta$ is a dimensionless measure of damping in a single junction (large
$\beta$ means low damping).  In our generalization, we take explicit 
account of the array lattice structure, so that a short-wavelength cutoff 
appears naturally in the resulting expression for the vortex friction 
coefficient.  The model gives semiquantitative agreement with our numerical 
experiments.  In particular, like the continuum model of Ref.\ 9 for square 
arrays, it accounts for the persistence of this friction coefficient even at 
high values of the shunt resistance $R$.  The present model 
is, in principle, applicable to dissipation from an arbitrary density of 
vortices moving with arbitrary velocities, and to losses produced by 
a.\ c.\ external currents.  However, we test it here only
for single vortex motion.

The remainder of this paper organized as follows. 
In Section II, we describe our calculational model and numerical
method.  Section III reports the results of our critical current 
calculations in triangular arrays both with and without vortices.
Section IV describes in greater detail our numerical results in the flux 
flow regime, and Sections V and VI summarize the vortex friction
model, and our investigation of possible ballistic vortex motion.
A brief discussion follows in Section VII.

\section{Model}

The relevant geometry of our triangular array is shown in Fig.\ 1.
Within the array interior of the array, each superconducting grain is
connected to its six nearest neighbors via Josephson coupling.   The 
boundary conditions involve fixed external current injection.  
We consider two 
directions of current injection, as illustrated in the Fig.\ 1,
the so-called [10\={1}] direction, and the [2\={1}\={1}] direction$^{20}$.

We describe the dynamics within the RCSJ model at zero temperature, 
as previously described for square arrays$^{14,21}$.
In this model, the current through each junction is the sum 
of three terms: a charge flow through an effective intergrain capacitance,
a current through a shunt resistance, and a Josephson supercurrent.
We assume that the supercurrent is sufficiently
weak that we can discard the induced screening current.

With these assumptions, the current between grains $i$ and $j$ is:
\begin{equation}
  I_{ij} = C_{ij}\frac{d}{dt}V_{ij} + \frac{V_{ij}}{R_{ij}}
         + I_{c;ij}\sin(\phi_i - \phi_j - A_{ij}).
\end{equation}
Here $I_{ij}$ is the total current from grain $i$ to grain $j$;
$C_{ij}$ and $R_{ij}$ are the shunt capacitance and shunt resistance
between grain $i$ and grain $j$; $I_{c;ij}$ is the critical current of the 
Josephson junction between grain $i$ and grain $j$; $\phi_i$ is the phase 
of the order parameter on grain $i$.  $V_{ij}$ and $A_{ij}$ are the voltage
difference and magnetic gauge phase factor between grain $i$ and grain $j$,
defined by
\begin{equation}
   V_{ij} \equiv V_i - V_j = \frac{\hbar}{2e}\frac{d}{dt} (\phi_i - \phi_j),
\end{equation}
and
\[ A_{ij}=\frac{2\pi}{\Phi_0}\int_{\bf{x}_i}^{\bf{x}_j}\bf{A}\cdot\bf{dl}, \]
where $\Phi_0 \equiv hc/(2e)$ is the flux quantum, $\bf{A}$
is the vector potential of 
the applied magnetic field and $\bf{x}_i$ is the position of the center of 
grain $i$.   In the present paper, we include only the intergrain capacitance, 
discarding the capacitance between the grains and ground.
Current conservation at each grain is described by Kirchhoff's Law,
\begin{equation}
    \sum_jI_{ij} = I_{i;ext},
\end{equation}
where $I_{i;ext}$ is the external current fed into grain $i$. 
We assume that
all the capacitances, critical currents, and shunt resistances
have unique values $C$, $I_c$, and $R$.
Finally, the use of classical equations of 
motion implies the assumption that quantum effects$^{22}$ arising from the 
noncommutativity of charge and phase variables can be neglected.

Our boundary conditions are shown in Fig.\ 1.  
In the direction of current 
injection, we introduce a uniform current $I_{i;ext}=I$ into each boundary
grain along one edge, and extract the uniform current from each boundary
grain on the other edge.  In the transverse direction, we use periodic
boundary conditions, as in our previous work$^{14}$.   The
gauge factor $A_{ij}$ satisfies
\[   \sum_{plaquette} A_{ij} = 2\pi \frac{BS}{\Phi_0} = 2\pi f,   \]
where $B$ is the magnetic field strength; $S$ is the area of each
triangular plaquette; $\Phi_0$ is the flux quantum. $f$ is the so-called
frustration.  Note that the primitive cell of a triangular array, 
consists of {\em two} adjacent triangular plaquettes.

The coupled equations (3), (4) and (5) are solved as described 
previously$^{14}$.  We use a fourth-order
Runge-Kutta algorithm with time step $\Delta t$, where $\Delta t$ ranges
from $0.01t_0$ to $0.05t_0$ ($t_0=\hbar/(2eRI_c)$ is a characteristic damping 
time), depending on the desired precision of calculation.  Further details 
may be found in Ref.\ 14.

\section{Critical Current and Vortex Depinning Current}

When $f=0$, with d.\ c.\ bias current injected in the [10\={1}] direction,
we find a critical current of exactly $2I_c$.  
This value can be easily understood, since in this case no current passes 
through the junctions perpendicular to the bias current, so that the entire 
array behaves much like a single junction.  With
[2\={1}\={1}] current injection, the critical current is found
to be approximately $1.76I_c$.  This value 
can be understood by considering the single triangular plaquette 
shown in Fig.\ 2. For such an arrangement, the injected current $I$ is 
related to the phase difference $\phi$ by $I/I_c=\sin \phi + \sin (2\phi)$. 
The right hand side cannot exceed $(I/I_c)_{max} = 1.7602$, 
which corresponds to our numerically obtained array value.  We have checked 
the phase configuration for each grain in the array in this geometry, and
find that it decomposes exactly into unit cells of this type.

Next, we discuss the dynamical response of an array under d.\ c.\ bias and
in the presence of a single vortex.  Such a vortex can be introduced by
considering an $N \times N$ array containing $2N^2$ triangular
plaquettes, and a flux $f=1/(2N^2)$ ($N$ being the number of junctions 
spanning the array in one direction, as in Fig.\ 1).
Table 1 lists the critical currents $I_d$ for $f = 1/(2N^2)$ 
as a function of $N$ for [10\={1}] current injection.
This critical current appears to be independent of $\beta$, at least
in the range $0 \leq \beta \leq 1000$.  The critical currents 
are extracted from an $I-\langle V\rangle$ plot, such as
shown in Fig.\ 3 for [10\={1}] current injection at 
$\beta= 0$ and $\beta = 10$, and Fig.\ 4 for [2\={1}\={1}] direction at 
$\beta = 0$. Extrapolating by eye a plot of $I_d$ versus $1/N$ towards 
$N = \infty$ at $f = 1/(2N^2)$, we estimate a critical current of about
$0.042I_c$ for bias current injected in the [10\={1}] direction.
In the [2\={1}\={1}] direction, for $N = 8$, we estimate a value of about
$0.041I_c$, possibly dependent on the initial phase configuration.
In both cases, our calculated critical currents are in reasonable agreement
with those obtained by Bobbert$^7$, using a piece-wise linear function to 
approximate the sinusoidal coupling function.

The array critical current at field $f = 1/(2N^2)$ can be interpreted as 
the depinning current of a single 
vortex.  It is of interest to compare our calculated value 
with the energy barrier for depinning.  This energy barrier was calculated
by Lobb {\it et al},$^{16}$, who used static methods to obtain a value
of about $0.043\hbar I_c/(2e)$.  This represents the energy which must be 
overcome in order to move a vortex from the center of one triangular 
plaquette to the center of an adjacent plaquette.

To make this comparison, and also to account for the apparent isotropy of
the vortex depinning current, we have used a simple model for the vortex 
potential $U({\bf r}) [{\bf r} \equiv (x, y)]$. Since $U(x, y)$ must have
the array periodicity, we express it as a Fourier series involving only 
Fourier components from the reciprocal lattice. The simplest approximation
consistent with the point symmetry of the lattice is to 
include only the smallest-magnitude Fourier components, i.\ e.
\begin{equation}
U({\bf r})=U_0+U_1\sum_{\bf K}\cos(\bf K \cdot \bf r)
\end{equation} 
where we take $U_1 > 0$ and $\bf K$ is one of the six smallest reciprocal 
lattice vectors.  (${\bf r}=0$ is interpreted as a grain center, and 
hence a maximum in the potential.)  The potential barrier for vortex 
depinning in this model is readily shown to be just $U_1$.

To estimate the critical current in this picture, we add to the vortex
potential energy a term $|\Phi_0{\bf J} \times {\bf r}|/c$, where ${\bf J}$
is the external current density.  This term corresponds to the Magnus force
${\bf J}\times \hat{{\bf z}}\Phi_0/c$ on a single vortex.  When this term 
is included, we find numerically that the barrier for vortex motion between
adjacent triangular plaquettes disappears at 
$a\hbar J/(2e) \approx 0.984 U_1$ for current injected in the [10\={1}] 
direction.  Taking $U_1 = 0.043\hbar I_c/(2e)$ from the results of Lobb 
{\it et al}$^{16}$, 
and using $J = I/a$ for [10\={1}] current injection, we see that 
our calculated vortex depinning current of $0.042I_c$ is in excellent 
agreement with the static results.  This calculation also suggests that 
expression (6) is a reasonable approximation for the vortex potential.

\section{Single Vortex Motion: Numerical Results}

As shown in Fig.\ 3 and Fig.\ 4, the $I-\langle V\rangle$ characteristics
display a long low voltage tail at currents above the
vortex depinning current.  This current regime is approximately Ohmic.
Since the $I-\langle V\rangle$ 
characteristics in this region can be understood in terms of single vortex 
motion in the array, this region is often 
called the flux flow regime$^{7-9}$.

At $\beta = 0$, the flux flow regime extends to about $1.9I_c$ and $1.5I_c$
for current biased in the [10\={1}] and [2\={1}\={1}] directions.
Both of these values are close to the critical current 
of the array at $f=0$.   Above these currents, the whole array is depinned 
and the vortex picture is not applicable. 
At sufficiently large values of $\beta$, with the current injected
at [10\={1}] direction, the flux flow regime terminates at lower currents,
where there is a ``row switching'' event$^{15, 17}$ rather than
the depinning of the entire array -- that is, one or more rows of junctions 
parallel to the direction of current injection switch from the supercurrent 
state to the resistively dissipative state.  This occurs, for example,
near $I = 1.5I_c$ at $\beta = 10$ in $8\times 8$ array.  Above this 
row-switching threshold, the picture of ohmic resistance by flux flow of 
vortices is no longer valid.  At yet higher $\beta$ values, 
the flux flow regime terminates at smaller currents, and 
there may be more than one
row switching event before the entire array is depinned.
At $\beta = 100$ in $8\times 8$ array, for example, we find two
row switching events in our calculations.   As in Ref.\ 9, 
we also find the staircase-like structure in the $I-\langle V\rangle$ curve
in the flux flow region, which may arise from the interaction of the vortex 
with its image neighboring vortices generated by the periodic boundary 
conditions.

Figs.\ 5(a) and 5(b) show the time-dependent space-averaged
voltage drop $V(t)$ across the array -- that is, the difference between 
the average voltage on the line of grains where the current is injected, 
and the line from which it is extracted -- at two current values in the 
flux-flow regime.  In both cases, $\beta = 0$ 
and the current is injected in the [10\={1}]
direction.  $V(t)$ is characterized by periodic sharp peaks which 
resemble the time-dependent voltage of a single junction, the frequency of 
which increases with increasing bias current.

If the single vortex picture is correct, the spike frequency $\nu_v$ is 
related to $\langle V\rangle$. The period of oscillation should correspond
to the motion of a vortex by one unit cell.   Furthermore, a complete vortex 
circuit around the lattice should produce a phase
change of $2\pi$ across the array.  Then using the 
Josephson relation, we obtain
\begin{equation}
  \langle V\rangle =\frac{\hbar}{2e}\langle\frac{d}{dt}\phi\rangle=
  \frac{\hbar}{2e}\frac{2\pi}{T},
\end{equation}
where $T$ is the period for one complete vortex circuit.
For an $N\times N$ array, $T = N/\nu_v$.
Our numerical results are in excellent agreement with this
relation, thus confirming the vortex motion picture in the flux flow regime.

By examining the time-dependent voltage of each single junction, one can 
also deduce the actual vortex path in the array.  This path is displayed in
Fig.\ 5(c) for an $8\times 8$ array with bias current in the
[10\={1}] direction.  We find that this path is independent of current
magnitude in the [10\={1}] direction. As shown in the figure, it is a straight
trajectory through the middle of the array.

The behavior of vortices is more complex when current is applied in the 
[2\={1}\={1}] direction. In this case, the flux flow regime in an $8\times 8$
overdamped array consists of
three distinct subregions with different slopes as shown in Fig.\ 4.
A typical voltage trace $V(t)$ from each subregion
is shown in Figs.\ 6(a), 6(b), and 6(c).  
The relationship between spike frequency and 
time-averaged voltage still holds, implying that the picture of vortex 
motion is still correct in this direction.  However, the vortex path is
different in each of the three subregions.  These paths, as deduced from
the time-dependent voltages of each junction, are shown in Figs.\ 6(d),
6(e), and 6(f).

At sufficiently low bias current in [10\={1}] direction, $V(t)$ shows a 
double-peaked structure.  We believe that this structure originates in the 
special geometry of the triangular lattice, in which each primitive cell has
two triangular plaquettes which are inequivalent.
When a bias current is applied in the [10\={1}] direction, the vortex will
pass alternately through each of these plaquettes, somehow producing a 
double-peaked structure in $V(t)$.  As the bias current increases, 
the double-peaked structure in $V(t)$ seems to disappear. 
However, the time-dependent voltage of each individual 
junction still exhibits a double-peaked structure.  The disappearance 
of the double-peak structure in $V(t)$ is therefore due to space-averaging.
We conclude that our simple vortex potential is qualitatively
correct even at higher bias current.   Of course, above 
the array depinning current ($I= 2I_c$ for this direction), 
the vortex picture breaks down and $V(t)$ shows no simple 
behavior, just as was found previously for square arrays$^{9, 14}$.

\section{Flux Flow Resistivity and Vortex Friction Coefficient}

As noted above, in the flux flow regime, the Josephson network behaves 
approximately ohmically. In this ohmic regime, 
we can define a {\em vortex friction coefficient} $\eta$ by 
equating the driving force
$J\Phi_0/c$ to the frictional drag force $\eta v$, where $v$ is the vortex 
velocity.  This gives (assuming current injected in the [10\={1}] direction)
\begin{equation}
   \eta v\equiv \frac{2\pi}{a}\frac{\hbar}{2e}I,
\end{equation}
where $I = aJ$ is the current injected into one node.

This vortex friction coefficient can be estimated in a simple way by equating
the frictional losses to the power dissipated in the shunt 
resistances when the vortex moves with constant velocity.  The result of 
this procedure for a square array is$^3$
\begin{equation}
  (\eta_0)_{sq}=\left(\frac{\hbar}{2e}\right)^2\left(\frac{2\pi}{a}\right)^2
  \frac{1}{2R},  
\end{equation}
and for a triangular array (cf.\ Appendix A), 
\begin{equation}
  (\eta_0)_{tri}=2\left(\frac{\hbar}{2e}\right)^2
   \left(\frac{2\pi}{a}\right)^2\frac{1}{2R}.
\end{equation}
$\eta$ can also be obtained directly from the calculated
$I-\langle V\rangle$ characteristics (cf.\ Appendix B).  Our numerical 
results for different values of $\beta$ in $8\times 8$ arrays are shown in 
Table 2. This Table shows that $\eta$ varies approximately as 
$\beta^{1/2}$ at large $\beta$ values, and differs considerably from the
result (10).  Indeed, at sufficiently large values
of $\beta$, the friction coefficient actually appears to be independent of
the shunt resistance $R$.  A similar result was also found previously
in square arrays.$^{7-9}$ 

A more accurate theoretical calculation of the frictional damping requires 
taking account of the loss of vortex energy to ``spin-wave-like'' excitations
in the array.   A continuum theory of this kind has been proposed by
Geigenm\"{u}ller {\it et al}$^{9}$.   In Appendix C, we
give a detailed, quasi-analytical theory for the friction coefficient,
based on the loss of energy from a vortex moving with velocity ${\bf v}$
to spin wave modes.  The model is more general than that of Ref.\ 9, in that 
it takes explicit account of the lattice structure of the array, so that a 
short-wavelength cutoff appears naturally, and 
allows for an arbitrary external current source to excite the spin wave modes
(for example, one arising from a high density of vortices). 
The final result for $\eta$ in a square array is 
\begin{equation}
 \left(\frac{\eta}{(\eta_0)_{sq}}\right) = \frac{1}{2\pi^2}I^{\prime},
\end{equation}
and in a triangular array
\begin{equation}
   \left(\frac{\eta}{(\eta_0)_{tri}}\right) = \frac{\sqrt{3}}{8\pi^2}
   I^{\prime}, 
\end{equation}
where the dimensionless integral $I^{\prime}$ is given in Appendix C.

It is sometimes useful to transform $\eta$ into an 
analogous expression for array resistivity.  Using the force balance 
equation (8), and noting that the electric field has magnitude 
$E=2\pi(\hbar/(2e))n_vv$, where $n_v=4f/(\sqrt{3}a^2)$
is the number of vortices per unit area, we obtain Ohm's Law in the form 
\[    {\bf E} = \rho {\bf J},       \]
where the array resistivity is (after considerable simplification)
\begin{equation}
\rho = \frac{32\pi^2fR}{3I^{\prime}}.
\end{equation}

Table 3 shows the friction coefficient as calculated from the model of 
Appendix C.  As can be seen, the resulting coefficient is strongly 
dependent upon $\beta$, in agreement with our numerical ``experiment.''  
If, for example, we choose the scaled vortex velocity
$\tilde{v}=1.0$ (as defined in Eq.\ (32)), 
our analytic expression for $\eta$ is well
approximated in square array by the simple expression
\begin{equation}
  \eta \approx 0.89\beta^{1/2}(\eta_0)_{sq},
\end{equation}
and in a triangular array by
\begin{equation}
  \eta \approx 0.34\beta^{1/2}(\eta_0)_{tri}.
\end{equation}
The $\beta^{1/2}$ trend is consistent with the results of numerical 
calculations in Ref.\ 9
for square arrays and in the present work for triangular arrays,
although the numerical coefficient may differ by as much as a factor of
two.  The trend is also consistent with the experimental results of Ref.\ 8.
Note that $\tilde{v} = 1$ is a reasonable velocity to consider in this 
comparison, because larger velocities tend to trigger row-switching events in
underdamped arrays.$^{7-10,15,23}$

Table 3 also lists the variation of $\eta$ with $\tilde{v}$ at several 
values of $\beta$.  Evidently, $\eta$ is nearly independent of
$\tilde{v}$ at large $\tilde{v}$ but goes to zero below a threshold value 
of order $\tilde{v} = 0.2$, where the damped pole in the integral (31) moves 
outside the first Brillouin zone of the triangular lattice.  Once again, 
this agrees with the results found in the continuum theory of Ref.\ 9 for 
a square lattice.  The results also agree quite well with our numerical 
results, as shown by a comparison of Tables 2 and 3(b).

\section{Absence of Ballistic Vortex Motion}

A rather surprising result of our simulations is the absence of ballistic 
vortex motion.  Such motion might have been expected, at least in the 
highly-underdamped regime; and there have been some experimental 
reports of such behavior$^{4,6}$.  In order to check for ballistic motion,
we apply a large bias current to the array in the flux flow regime, 
so that the vortex acquires a high initial velocity, then we
turn off the bias current.  Since the effective mass of the vortex is
presumably large in the high-$\beta$ regime, the vortex might be 
expected to move several
lattice spacings because of its large initial kinetic energy, even after 
the driving current is removed.  But from the calculated time-dependent 
voltage of the individual junctions, we find that the vortex travels at
most through one primitive cell after the bias current is shut off, 
whatever its initial velocity, for all values of $\beta$ considered
(0 $\leq \beta \leq 1000$).  This is consistent with previous 
calculations$^{7}$. 

It is of some interest to compare our results with those of Ref. 10.  This 
paper considers the triangular array in a continuum approximation and 
concludes that a narrow vortex velocity window exists where ballistic 
motionis possible.  It is suggested 
that this window extends from the vortex depinning current to roughly twice 
that current.  We can envision two possible reasons why we do not see this 
ballistic regime in our own calculations.  The first is that the depinning 
current is rather dependent on lattice size, typically being larger for the 
smaller lattices.  For our size regime, this may narrow the window nearly 
to zero.  In addition, the vortex velocity is not constant just above 
the depinning current, but instead is quite time-dependent, because of the 
periodic pinning potential.  This time dependence is not considered in the 
model of ref. 10, which assumes a constant vortex velocity in estimating 
the width of the ballistic ``window.''  Thus, the 
effects of this time-dependence could possibly also suppress this window.  

In view of these results, the explanation for the ballistic motion which is 
observed in experiments seems unclear.  No numerical calculation has yet 
found such motion from classical equations.  Conceivably, the ballistic 
regime arises when quantum effects reduce the vortex friction 
coefficient below classical predictions, but this remains to be proven.

\vspace{0.6in}

\section{Discussion and Conclusions}

We have simulated the dynamical response of triangular Josephson junction 
arrays, using a classical model of resistively- and capacitively-shunted 
Josephson junctions described by coupled second-order 
differential equations.  In the flux flow regime, the dynamical response of 
the network, including the time-dependent voltage, 
is well described in terms of vortex degrees of freedom.  The 
vortices, however, experience higher viscous damping than predicted on the 
basis of a simplified model, and 
in apparent contrast to experiment do not exhibit 
ballistic motion at any bias current we have investigated.  The damping is 
reasonably well described, however, by a model which describes loss of
vortex energy to plasma (or ``phase wave'')
oscillations in the Josephson network. 

It is of interest to make a crude estimate of
the parameters where quantum corrections 
might need to be included in these calculations.  In a naive picture,
such corrections would start to matter when the characteristic charging 
energy $(2e)^2/C$ becomes comparable to the Josephson energy 
$\hbar I_c/(2e)$.  This condition gives
\begin{equation}
  CI_c \approx \frac{(2e)^3}{\hbar}.
\end{equation}
This can be translated into a condition on the lattice resistivity, using
Eqs.\ (12), (13), and (15), with the result 
\begin{equation}
  \rho \propto f\left(\frac{h}{(2e)^2}\right),
\end{equation}
where the constant of proportionality is approximately 1.1.  This result 
suggests that quantum corrections might become 
important when the resistance per 
square of this two-dimensional network is comparable to  the ``quantum of 
resistance'' $h/(2e)^2$ 
per unit frustration.  Of course, this naive estimate does not 
take account of such obvious corrections as vortex-vortex interactions.  
It is amusing to note that several groups 
have reported evidence (both experimental and theoretical$^{24}$) for 
a superconductor-to-insulator transition in quasi-two-dimensional 
superconductors in a magnetic field at a resistance per square of order
$h/(2e)^2$; this transition is generally attributed to disorder effects, 
and thus may be unrelated to the simple criterion for arrays mentioned 
above.  Thus a detailed calculation of quantum effects on vortex motion 
remains an important problem for future study.

\vspace{0.6in}

\section{Acknowledgments}
 
We gratefully acknowledge support by 
the National Science Foundation through Grant 
DMR 90-20994.  Our calculations were carried out, in part, on the
CRAY Y-MP 8/864 of the Ohio Supercomputer Center, with the help of a grant
of time which we gratefully acknowledge.

\newpage

\begin{center}
APPENDIX A: SIMPLE ESTIMATE OF VORTEX FRICTION COEFFICIENT
\end{center}

\vspace{0.3in}

In this Appendix, we present a simple estimation of the friction coefficient
for a single vortex moving in a triangular array.
If we assume that such a vortex moves from the center of one plaquette
to the center of a nearest-neighbor plaquette, it must cross one junction.
Since the change in phase difference 
is $2\pi /3$ when the vortex crosses the junction, 
the average voltage across the junction is
\[  \langle V\rangle = \frac{\hbar}{2e}\cdot\frac{2\pi /3}{\Delta t} = 
    \frac{\hbar}{2e}\cdot\frac{2\pi /3}{\frac{\sqrt{3}}{3}a/v} = 
    \frac{2}{\sqrt{3}}\cdot\frac{\hbar}{2e}\cdot\frac{\pi}{a}\cdot v,   \]
where $\Delta t$ is the time required for the vortex to move
from one plaquette center to the next, $a$ is the lattice constant, 
and $v$ is the time-averaged vortex velocity.

We next compute the effective frictional coefficient by equating the
resistively dissipated energy to that expected for a particle moving in a 
viscous medium.  Now  
the effective resistance between two nearest neighbor grains is defined as 
the voltage drop which is produced when a unit current is injected into 
one such grain and extracted from the other.  Since there are six nearest 
neighbors in a triangular lattice, the effective resistance in an infinite 
triangular array is 
$R/3$, where $R$ is the single-junction resistance.$^{25}$  
Equating the resistively dissipated power to the
frictional losses, we obtain
\[  \frac{1}{2}(\eta_0)_{tri}v^2=\frac{\langle V\rangle^2}{2(R/3)}
    =\frac{3}{2R}\cdot\frac{4}{3}\cdot (\frac{\hbar}{2e})^2\cdot
     (\frac{\pi}{a})^2\cdot v^2.   \]
This implies that the vortex frictional
coefficient in an infinite triangular array is
\begin{equation}
  (\eta_0)_{tri}=2\cdot (\frac{\hbar}{2e})^2\cdot (\frac{2\pi}{a})^2
    \cdot\frac{1}{2R} .
\end{equation}
A similar calculation for a square array gives$^{3, 5}$:
\begin{equation}
  (\eta_0)_{sq}= (\frac{\hbar}{2e})^2\cdot (\frac{2\pi}{a})^2
    \cdot\frac{1}{2R}. 
\end{equation}

\vspace{0.5in}

\begin{center}
APPENDIX B: EXTRACTION OF $\eta$ FROM $I-\langle V\rangle$ CHARACTERISTICS
\end{center}
\vspace{0.3in}

In the flux flow region, the time- and space-averaged voltage 
$\langle V\rangle$ across an $M\times N$ array is approximately proportional 
to the bias current $I$. 
We define a dimensionless proportionality coefficient $\gamma$ by
\[ \langle V\rangle = \gamma NRI, \]
where $N$ is the number of junctions along the direction of the bias 
current.   If we assume periodic transverse boundary conditions
and [10\={1}] current injection, a complete 
circuit of a vortex around the array produces a phase 
change of $2\pi$ across the array in the direction parallel to the current 
injection.  The Josephson relation then implies that
\[ \langle V\rangle = \frac{\hbar}{2e}\langle\frac{d\phi}{dt}\rangle
   = \frac{\hbar}{2e}\frac{2\pi}{Ma}v,            \]
where $a$ is the lattice constant and $v$ is the transverse vortex velocity.
Since the friction coefficient $\eta_{tri}$ is related to $v$
by Eq.\ (8), we obtain, on combining the above relations,
\begin{eqnarray}
  \eta_{tri} & = & \frac{2}{\gamma MN}\cdot \left(\frac{\hbar}{2e}\right)^2 
           \cdot\left(\frac{2\pi}{a}\right)^2\cdot \frac{1}{2R} \nonumber  \\
             & = & \frac{1}{\gamma MN} (\eta_0)_{tri}.
\end{eqnarray}
\vspace{0.3in}

\begin{center}
APPENDIX C: ANALYTICAL MODEL FOR VORTEX FRICTION COEFFICIENT
\end{center}

\vspace{0.3in}

We consider a d-dimensional periodic 
network of RCSJ's, assuming zero shunt capacitance to ground, and also 
assuming that the shunt resistance, 
junction critical current, and shunt capacitance all
vanish except between nearest neighbors, for which they take the 
values $R$, $I_c$, and $C$ respectively.

With these assumptions, the equations of motion for the phases may be 
written in the form
\begin{equation}
  \frac{\hbar}{2e}C\sum_j\ddot{\phi}_{ij}
  +\frac{\hbar}{2eR}\sum_j\dot{\phi}_{ij}
  +I_c\sum_j\sin(\phi_{ij})=I_{i;ext},
\end{equation}
where the sums run over the nearest neighbors to the grain $i$ and
$\phi_{ij} = \phi_i-\phi_j$.

We will calculate the losses produced by an externally applied current due 
to excitation of ``spin waves'', i.\ e.\ small-amplitude phase fluctuations.
Thus, we expand the sine-function as
$\sin(\phi_i-\phi_j) \approx \phi_i-\phi_j$,
and, within that assumption, calculate the losses coming from an arbitrary
$I_{i;ext}(t)$.  With the introduction of the Fourier transforms
$\phi({\bf k}, \omega)$ and
$I_{ext}({\bf k}, \omega)$,
we can transform (21) into the form
\begin{equation}
   \phi({\bf k},\omega)=\frac{I_{ext}({\bf k},\omega)/t({\bf k})}{I_c
    -\frac{i\hbar\omega}{2eR}-\frac{\hbar\omega^2C}{2e}}.
\end{equation}
Here
\[  t({\bf k}) = \sum_{nn}(1-\exp(i{\bf k}\cdot\bf{R})),     \]
the sum runs over the set of nearest neighbor lattice vectors 
${\bf R}$, and the
allowed ${\bf k}$ values run over the first Brillouin zone of the 
grain lattice.

The energy dissipation in the $ij^{th}$ bond in the time interval $[-T, T]$ is:
\[  \Delta E_{ij} = \int_{-T}^{+T}I_{ij}(t)V_{ij}(t)dt,   \]
or, in Fourier transform,
\begin{equation}
  \Delta E_{ij} = \int_{-\infty}^{\infty}d\omega
  \int_{-\infty}^{\infty}d\omega^{\prime}I_{ij}(\omega)
  V_{ij}^*(\omega^{\prime})A(\omega, \omega^{\prime}),
\end{equation}
where
\[  A(\omega,\omega^{\prime}) = \frac{1}{\pi}
    \frac{\sin[(\omega-\omega^{\prime})T]}{\omega-\omega^{\prime}}.   \]
The total energy loss $\Delta E_{tot} \equiv \sum_{<ij>}\Delta E_{ij}$.
Using the Josephson relation between phase and voltage, and the equations 
of motion (21) in the small phase difference approximation, and making the 
relevant Fourier transforms, we can finally express the total energy loss 
as
\begin{eqnarray}
  \Delta E_{tot} & = & \frac{\hbar}{2eN}\sum_{{\bf k}}\frac{1}{t({\bf k})}
  \int_{-\infty}^{\infty}d\omega\int_{-\infty}^{\infty}d\omega^{\prime} 
   \times  \nonumber \\ 
                 & \times &\Re \frac{i\omega^{\prime}I_{ext}({\bf k},\omega)
  I_{ext}^*({\bf k},\omega^{\prime})}{I_c+\frac{i\hbar\omega^{\prime}}{2eR}
  -\frac{\hbar\omega^{\prime 2}C}{2e}} A(\omega,\omega^{\prime}),
\end{eqnarray}
where $N$ is the number of grains in the lattice, and the sum runs over the 
first Brillouin zone.

This result is valid for an arbitrary external current source.  We now 
specialize to a vortex moving with velocity $\vec{v}$.
According to Geigenm\"{u}ller {\it et al}, such a vortex traveling
in the y direction has associated with it a charge density
\[  Q^{V}({\bf x}, t) = C\frac{\hbar}{2e}v\frac{\partial}{\partial x}2\pi
    \delta(x)\delta(y-vt).   \]
The corresponding space and time Fourier transform is:
\begin{equation}
  Q^{V}({\bf k},\omega) = (2\pi)^{3/2} C\frac{\hbar}{2e}v(-ik_x)
     \delta(\omega -vk_y).
\end{equation}
Of course, this charge density was derived for a vortex moving in a
{\em continuum} superconductor, and cannot be exactly correct for a 
superconducting array.  However, by imposing the additional requirement that
Q$^{V}({\bf k},\omega)$ vanish for {\bf k} outside the first Brillouin 
zone, we produce an approximate charge density which is properly discrete.

The external current corresponding to this charge distribution is:
\[  I_{ext}({\bf k},\omega) = -i\omega Q^V({\bf k}, \omega).  \]
We substitute this into Eq.\ (24) and convert the sum 
over ${\bf k}$ into an integral, with the help of
a factor $S/(4\pi^2)$, where $S$ is the area of the array. 
The integrand vanishes unless $\omega = \omega^{\prime}$.  
Next, we explicitly evaluate the real part, using the fact that 
$A(\omega,\omega) = \frac{T}{\pi}$, and dividing by $2T$.
The two integrals over frequency can be done immediately, since they 
involve delta-functions, and the final result for the time rate of energy 
loss from the vortex into the spin wave modes reduces to
\begin{equation}
  \frac{dE}{dt} = \eta v^2,
\end{equation}
where $\eta$ is an effective \underline{vortex friction coefficient}.
The form (26) properly corresponds to a frictional force of the form $-\eta v$,
since the rate of energy loss is the dot product of 
the force with the velocity.

The expression for $\eta$ can be written most compactly by using the 
dimensionless variables
$k_i^{\prime} = ak_i$, $i = x$ or $y$, where $a$ is the bond length.  We 
also introduce a dimensionless vortex velocity 
\begin{equation}
\tilde{v} = v/(a\omega_0),
\end{equation}
where $\omega_0 = \sqrt{2eI_c/\hbar C}$ is the Josephson plasma frequency.
After some algebra, we obtain Eqs.\ (11) and (12), respectively, 
for $\eta$ in a square array and a triangular array.
In both cases the dimensionless integral $I^{\prime}$ takes the form
\begin{equation}
  I^{\prime}(\tilde{v},\beta) = \int_{B. Z.}dk_x^{\prime}dk_y^{\prime}
  \left(\frac{(k_x^{\prime})^2(k_y^{\prime})^4}{[1/\tilde{v}^2
  -(k_y^{\prime})^2]^2+\frac{(k_y^{\prime})^2}{\tilde{v}^2\beta}}\right)
  \left(\frac{1}{t(k_x^{\prime},k_y^{\prime})}\right),
\end{equation}
and $(\eta_0)_{sq}$ and $(\eta_0)_{tri}$ 
are given by Eq.\ (9) and Eq.\ (10).  The
integral for both lattices 
runs over the scaled first Brillouin zone of the array
(defined by taking the bond length $a = 1$).

\newpage
\begin{center}
REFERENCES AND FOOTNOTES
\end{center}

\vspace{0.1in}
 
\begin{enumerate}

\item
For many references up to 1988, see, e.\ g., the articles in Physica
(Amsterdam) {\bf 152B}, pp.\ 1-302 (1988).  For some recent reviews, see, 
e.\ g.,
the articles in {\it Proceedings of the 2nd CTP Workshop on Statistical 
Physics: KT Transition and Superconducting Arrays}, edited by D. Kim, 
J. S. Chung, and M. Y. Choi (Min Eum Sa, Seoul, Korea, 1993).

\item
U. Eckern and A. Schmid, Phys.\ Rev.\ {\bf B39}, 6441 (1989).

\item
M. S. Rzchowski, S. P. Benz, M. Tinkham and C. J. Lobb,
Phys.\ Rev.\ {\bf B42}, 2041 (1990).

\item
H. S. J. van der Zant, F. C. Fritschy, T. P. Orlando and J. E. Mooij,
Physica {\bf B165-66}, 969 (1990).

\item
T. P. Orlando, J. E. Mooij and H. S. J. van der Zant,
Phys.\ Rev.\ {\bf B43}, 10218 (1991).

\item
H. S. J. van der Zant, F. C. Fritschy, T. P. Orlando and J. E. Mooij,
Europhys.\ Lett.\ {\bf 18}, 343 (1992).
 
\item
P. A. Bobbert, Phys.\ Rev.\ {\bf B45}, 7540 (1992).

\item
H. S. J. van der Zant, F. C. Fritschy, T. P. Orlando and J. E. Mooij,
Phys.\ Rev.\ {\bf B47}, 295 (1993).

\item
U. Geigenm\"{u}ller, C. J. Lobb, C. B. Whan, Phys.\ Rev.\ {\bf B47}, 
348 (1993).
  
\item U. Eckern and E. B. Sonin, Phys.\ Rev.\ {\bf B47}, 505 (1993).

\item B. J. Van Wees, Phys.\ Rev.\ Lett.\ {\bf 65}, 255 (1990).

\item T. P. Orlando and K. A. Delin, Phys.\ Rev.\ {\bf B43}, 8717 (1991).

\item R. Th\'{e}ron, J.-B. Simond, Ch.\ Leemann, H. Beck, P. Martinoli,
and P. Minnhagen, Phys.\ Rev.\ Lett.\ {\bf 71}, 1246 (1993).

\item
Wenbin Yu, K. H. Lee and D. Stroud, Phys.\ Rev.\ {\bf B 47}, 5906 (1993).

\item
Wenbin Yu and D. Stroud, Phys.\ Rev.\ {\bf B 46}, 14005 (1992).

\item
C. J. Lobb, D. W. Abraham and M. Tinkham, Phys.\ Rev.\ {\bf B27}, 150 (1983).

\item
H. S. J. van der Zant, C. J. Muller, L. J. Geerligs, C. J. P. M. Harmans
and J. E. Mooij, Phys.\ Rev.\ {\bf B38}, 5154 (1988).

\item
T. S. Tighe, A. T. Johnson and M. Tinkham, Phys.\ Rev.\ {\bf B44},
10286 (1991).

\item D. E. McCumber, J. Appl.\ Phys.\ {\bf 39}, 3113 (1968);
W. C. Stewart, Appl.\ Phys.\ Lett.\ {\bf 22}, 277 (1968).

\item
L. L. Sohn, M. S. Rzchowski, J. U. Free, M. Tinkham, and C. J. Lobb,
Phys.\ Rev.\ {\bf B45}, 3003 (1992).

\item
There have been, by now, a very large number of dynamical calculations
based on similar sets of coupled equations.  Some representative 
calculations are: 
K. K. Mon and S. Teitel, Phys.\ Rev.\ Lett.\ {\bf 62}, 673 (1989); 
A. Falo {\it et al},     Phys.\ Rev.\ {\bf B41}, 10983 (1990); 
T. C. Halsey,            Phys.\ Rev.\ {\bf B41}, 11634 (1990); 
W. Xia and P. L. Leath,  Phys.\ Rev.\ Lett.\ {\bf 63}, 1428 (1989); 
H. Eikmans and J. E. van Himbergen, Phys.\ Rev.\ {\bf B41}, 8927 (1990); 
L. L. Sohn {\it et al},  Phys.\ Rev.\ {\bf B44}, 925 (1991); 
D. Dominguez {\it et al}, Phys.\ Rev.\ Lett.\ {\bf 67}, 2367 (1991);
M. Octavio {\it et al},  Phys.\ Rev.\ {\bf B44}, 4601 (1991); 
K. Y. Tsang {\it et al}, Phys.\ Rev.\ Lett.\ {\bf 66}, 1094 (1991); 
R. Bhagavatula {\it et al}, Phys.\ Rev.\ {\bf B45}, 4774 (1992); 
S. R. Shenoy, J. Phys.\ {\bf C18}, 5163 (1985).

\item 
Quantum effects have been discussed by many authors, starting with
P. W. Anderson, in {\em Lectures on the Many Body Problem}, edited by
E. R. Caianello (Academic Press, New York 1964), Vol II;
B. Abeles, Phys.\ Rev.\ {\bf B15}, 2828 (1977); and
E. Simanek, Solid State Commun.\ {\bf 31}, 419 (1979).  For further 
references, see, e.\ g., G. Sch\"{o}n in Ref.\ 1, and references therein.

\item This criterion for the excitation of spin waves
was pointed out long ago, in the context of a 
continuum model, by K. Nakajima and Y. Sawada, 
J. Appl.\ Phys.\ {\bf 52}, 5732 (1981).

\item M. P. A. Fisher, Phys.\ Rev.\ Lett.\ {\bf 65}, 923 (1990); 
A. F. Hebard and M. A. Paalanen, Phys.\ Rev.\ {\bf 65}, 927 (1990); 
E. S. S\/{o}renson {\it et al}, Phys.\ Rev.\ Lett.\ {\bf 69}, 828 (1992). 

\item S. Kirkpatrick, Rev.\ Mod.\ Phys.\ {\bf 45}, 574 (1973).

\end{enumerate}
 
\newpage
\begin{center}
FIGURE CAPTIONS
\end{center}
\vspace{0.1in}

\begin{enumerate}
 
\item  
Schematic diagram of an $8\times 8$ triangular Josephson junction array. 
Each intersection
represents a superconducting grain, which is connected to its six nearest
neighbors by Josephson coupling.  (a) and (b) correspond respectively to
[10\={1}] and [2\={1}\={1}] current injection direction.   Free boundary 
conditions are used in the direction of current 
injection, while periodic boundary conditions are used in the transverse 
direction.

\item
Schematic illustration of a triangular plaquette of Josephson junctions
at zero magnetic field, subjected to an injected current $I$ as shown.
The critical current for this arrangement is $1.7602I_c$.

\item
$I-\langle V\rangle$ characteristics for $8\times 8$ triangular arrays at 
$f=1/128$ at two different $\beta$ values with current injected in
the [10\={1}] direction: (a) $\beta = 0$; (b) $\beta = 10$. The insets are
enlargements of the flux flow regime.

\item
$I-\langle V\rangle$ characteristics for overdamped $8\times 8$ triangular
arrays ($\beta = 0$) at $f=1/128$ with current injected in the 
[2\={1}\={1}] direction. The inset is the enlargement of the flux flow regime.

\item
Time dependent voltage traces and vortex motion path in the array for 
$8\times 8$ overdamped arrays ($\beta = 0$) at $f=1/128$ with current 
injected in the [10\={1}] direction for two different 
applied currents. $t_0 = \hbar / (2eRI_c)$ is a 
natural unit of time. The bias current and time-averaged voltages are
(a) $I/I_c=0.2$, $\langle V\rangle /NRI_c=2.39\times 10^{-3}$;  (b) 
$I/I_c=1.0$, $\langle V\rangle /NRI_c=1.91\times 10^{-2}$.
(c) Path of vortex motion in this array; the disks represent successive
positions of the vortex in the array.

\item
Time dependent voltage traces and vortex trajectories in an 
$8\times 8$ overdamped array ($\beta = 0$) at $f=1/128$, for three different 
values of current applied in the [2\={1}\={1}] direction.
$t_0 = \hbar / (2eRI_c)$ is the natural unit of time.  The bias currents and 
time-averaged voltages are (a) $I/I_c=0.4, \langle V\rangle /NRI_c=4.298
\times 10^{-3}$; (b) $I/I_c=0.9, \langle V\rangle /NRI_c=1.123\times 10^{-2}$;
and (c) $I/I_c=1.3, \langle V\rangle /NRI_c=1.978\times 10^{-2}$.  (d), (e), 
and (f) show the vortex trajectories corresponding to (a), (b), and (c).
We draw the trajectories in a ``repeated lattice scheme'' in which the 
periodic boundary conditions are represented by repeating the 
$N\times N$ lattice.

\end{enumerate}

\newpage
\begin{center}
TABLE CAPTIONS
\end{center}
\vspace{0.1in}
\begin{enumerate}

\item
Numerical values of the critical current for an $N\times N$ triangular array
at $f=1/(2N^2)$, for [10\={1}] current injection.

\item
Numerical values of $\eta$ as a function of $\beta$ in an $8\times 8$
triangular array.   $\gamma$ is estimated from the numerical
$I-\langle V\rangle$ characteristics of the corresponding arrays.
$\eta$ is calculated from Eq.\ (20) in Appendix B, and is estimated by 
drawing a straight line by eye through the $I-\langle V\rangle$ 
characteristic in the 
flux-flow regime.

\item
Numerical values of $\eta$ as obtained from the semi-analytical theory of 
Appendix C at several values of $\beta$ and $\tilde{v}$ for square and 
triangular arrays.  The values are obtained by use of Eqs.\ (33) -- (35), 
carrying out the integral numerically.

\end{enumerate}

\begin{table}
 \begin{center}
     {\bf Table 1}
 \end{center}

 \begin{center}
  \begin{tabular}
    {||c|c|c|c|c||} \hline \hline
    $N$        &   8   &  12   &  16   &  24     \\ \hline
    $I_d/I_c$  & 0.090 & 0.063 & 0.054 & 0.048   \\ \hline \hline
  \end{tabular}
 \end{center}
\end{table}

\vspace{0.6in}

\begin{table}
 \begin{center}
     {\bf Table 2}
 \end{center}

 \begin{center}
  \begin{tabular}
    {||c|c|c||} \hline \hline
    $\beta$ &  $\gamma$            & $\eta/\eta_0$   \\ \hline \hline
        0   & $2.73\times 10^{-2}$ &   0.572         \\ \hline
        1   & $2.13\times 10^{-2}$ &   0.732         \\ \hline
       10   & $8.57\times 10^{-3}$ &   1.82          \\ \hline
       50   & $4.09\times 10^{-3}$ &   3.82          \\ \hline
      100   & $2.78\times 10^{-3}$ &   5.63          \\ \hline
      225   & $2.07\times 10^{-3}$ &   7.54          \\ \hline
      400   & $1.47\times 10^{-3}$ &   10.7          \\ \hline \hline
  \end{tabular}
 \end{center}
\end{table}

\newpage 

\begin{table}
 \begin{center}
     {\bf Table 3 (a)}
 \end{center}

 \begin{center}
  \begin{tabular}
    {||c|c|c|c|c|c|c|c||} \hline \hline
    \multicolumn{2}{||c|}{ }  &  \multicolumn{6}{c||}{$\beta$}  \\ \cline{3-8}
    \multicolumn{2}{||c|}{$\eta / (\eta_0)_{sq}$} & 1 & 10 & 50 & 100 & 225 &
                400 \\  \hline
           & 0.2 & 0.040 & 0.061 & 0.064 & 0.064 & 0.064 & 0.064 \\ \cline{2-8}
           & 0.5 & 0.67 & 3.16 & 7.67 & 11.0 & 16.7 & 22.4 \\ \cline{2-8}
$\tilde{v}$& 1.0 & 1.14 & 3.25 & 6.75 & 9.33 & 13.7 & 18.2 \\ \cline{2-8}
           & 2.0 & 1.38 & 2.82 & 5.10 & 6.83 & 10.1 & 14.1 \\ \hline \hline
  \end{tabular}
 \end{center}

\vspace{0.5in}
 \begin{center}
     {\bf Table 3 (b)}
 \end{center}

 \begin{center}
  \begin{tabular}
    {||c|c|c|c|c|c|c|c||} \hline \hline
    \multicolumn{2}{||c|}{  }  & \multicolumn{6}{c||}{$\beta$}  \\ \cline{3-8}
    \multicolumn{2}{||c|}{$\eta / (\eta_0)_{tri}$} & 1 & 10 & 50 & 100 & 225 &
                400 \\  \hline
           & 0.2 & 0.014 & 0.025 & 0.027 & 0.027 & 0.028 & 0.028 \\ \cline{2-8}
           & 0.5 & 0.20 & 0.94 & 2.32 & 3.36 & 5.13 & 6.91 \\ \cline{2-8}
$\tilde{v}$& 1.0 & 0.38 & 1.19 & 2.55 & 3.56 & 5.25 & 6.84 \\ \cline{2-8}
           & 2.0 & 0.48 & 1.06 & 1.98 & 2.66 & 3.86 & 5.06 \\ \hline \hline
  \end{tabular}
 \end{center}
\end{table}
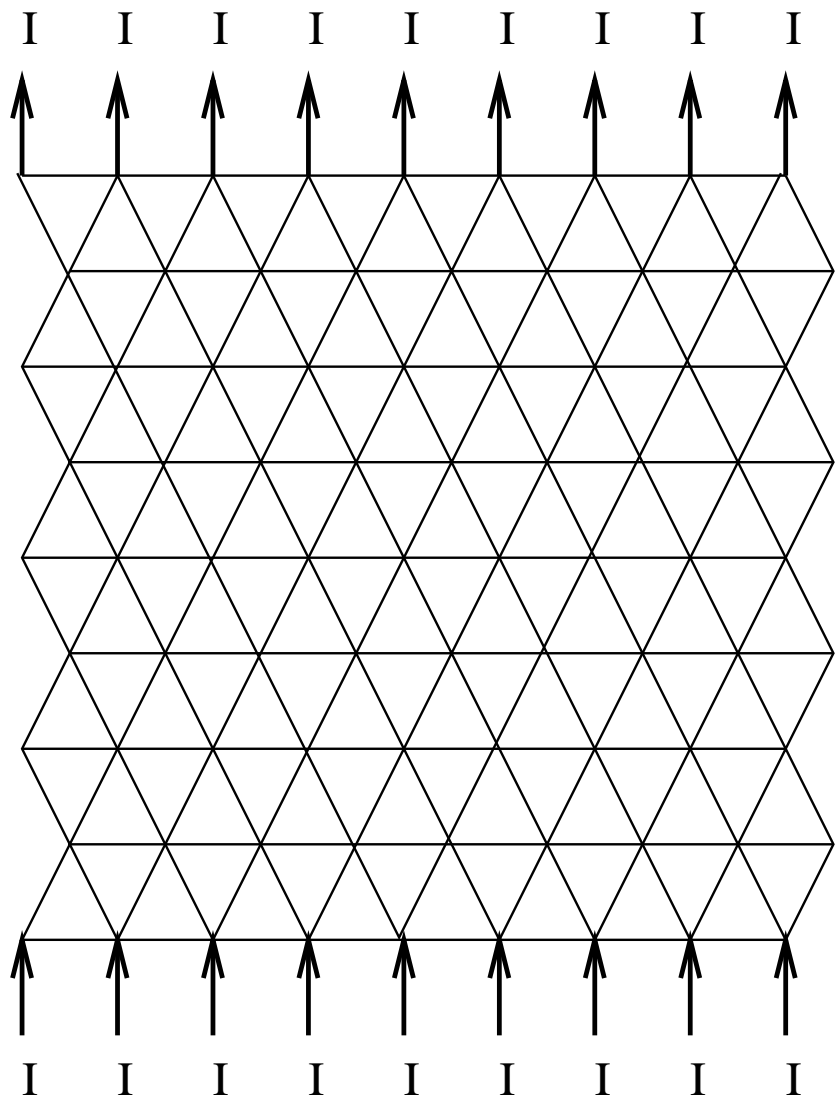
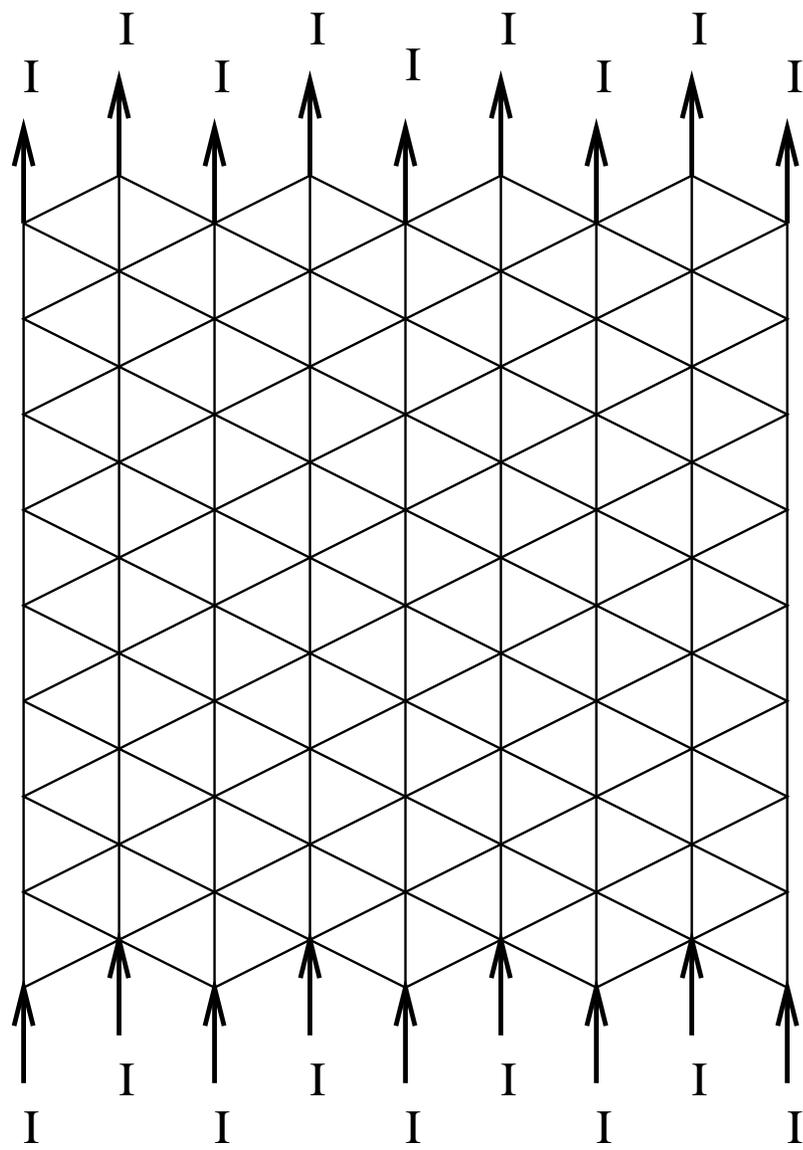
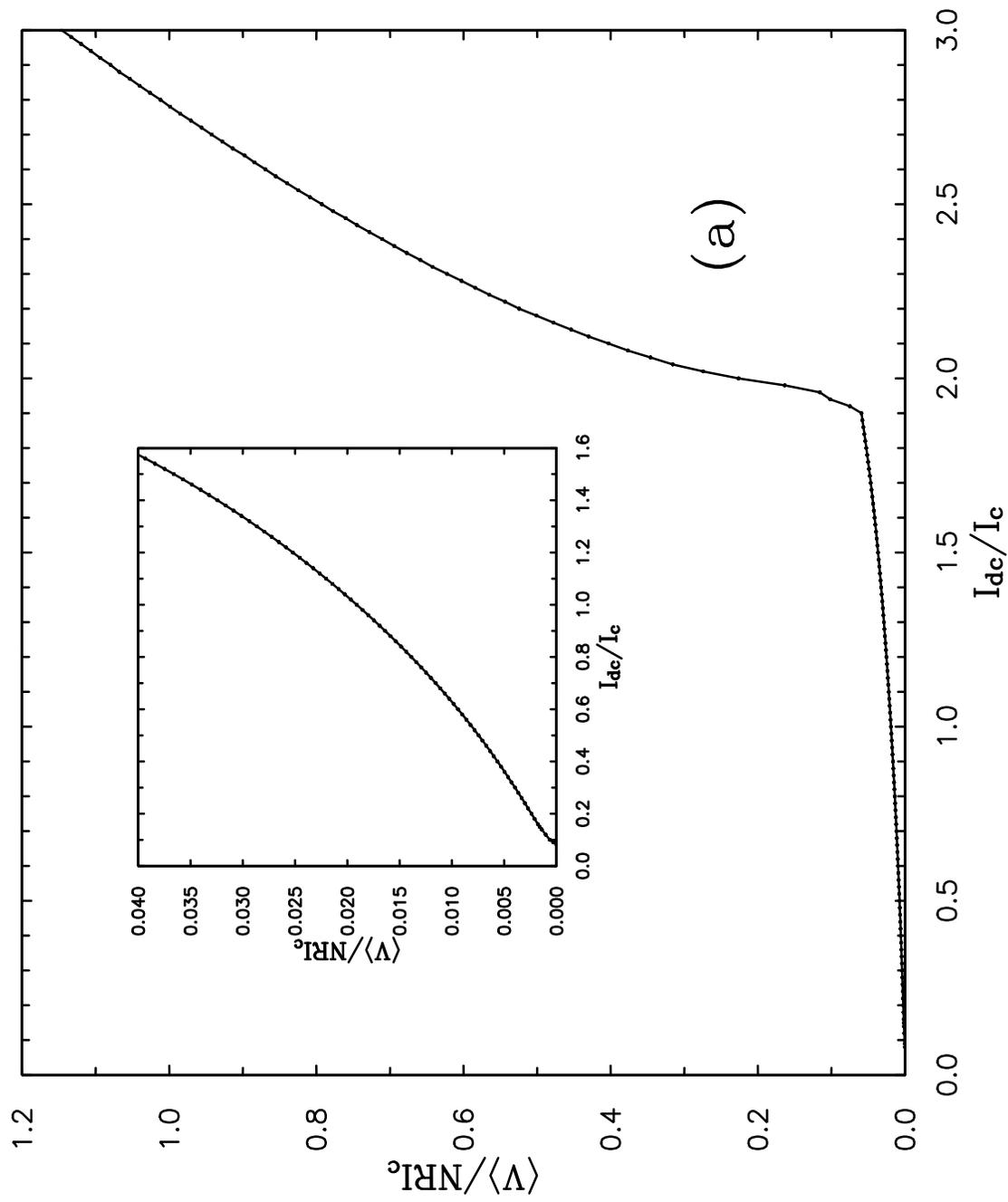
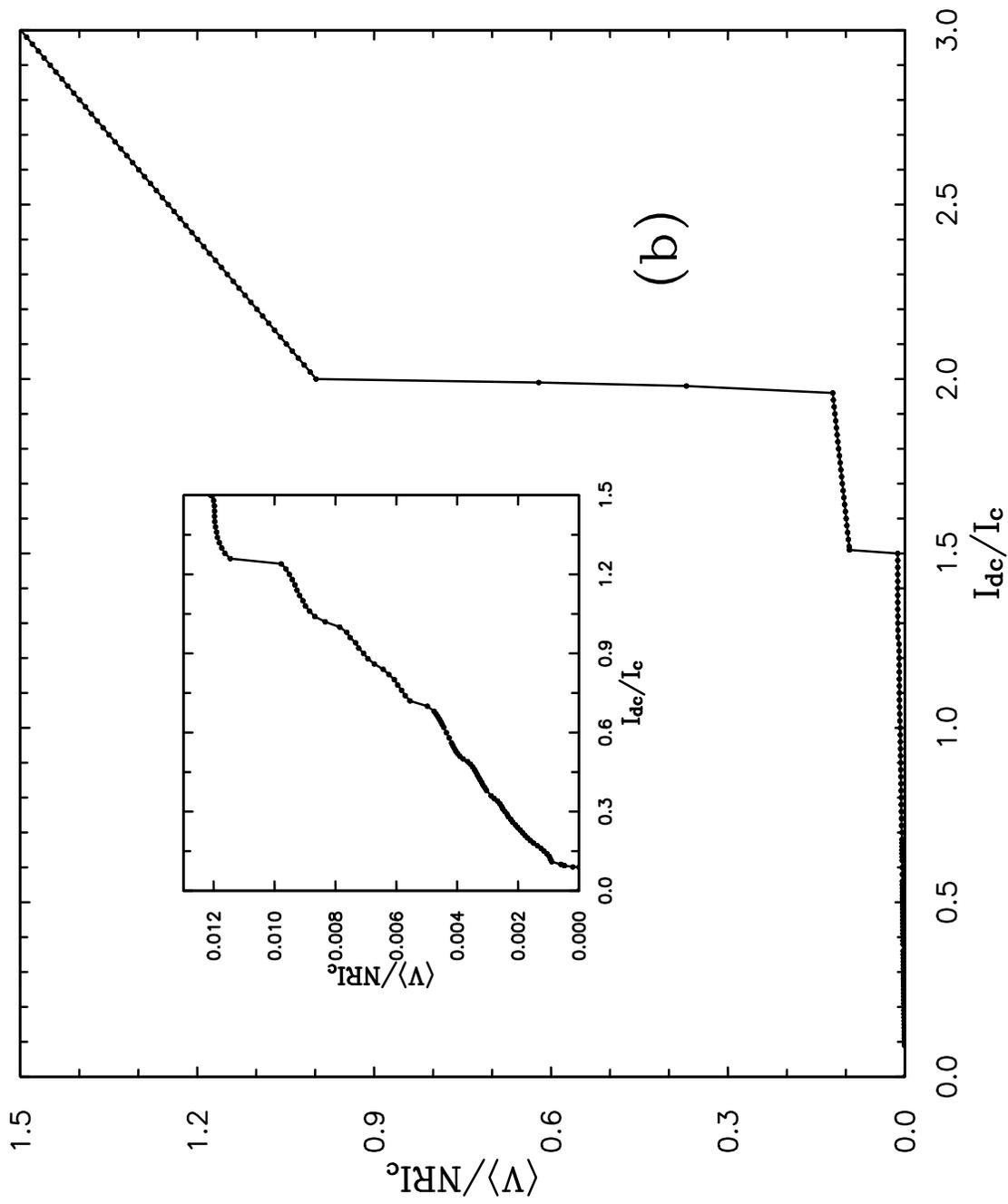
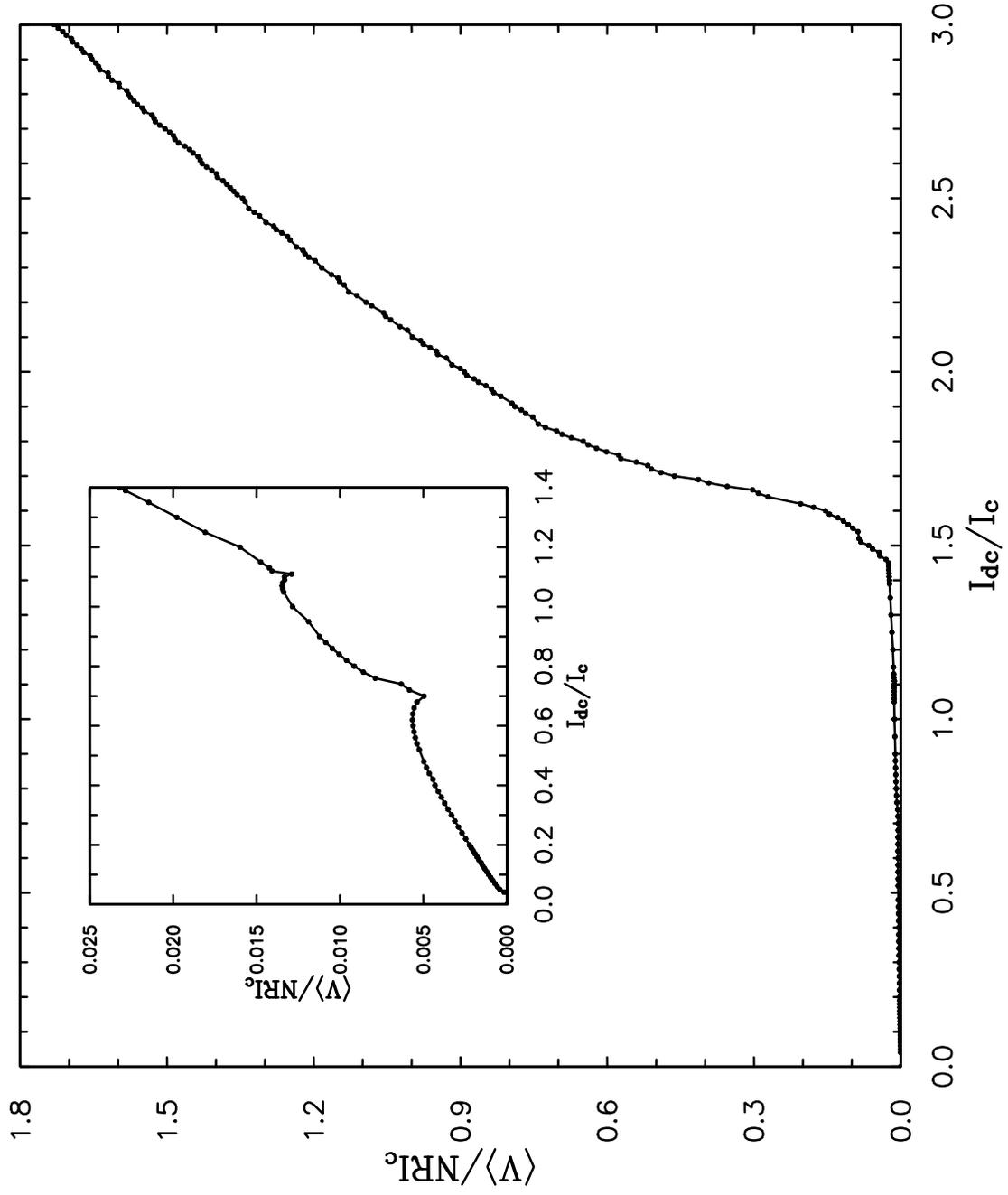
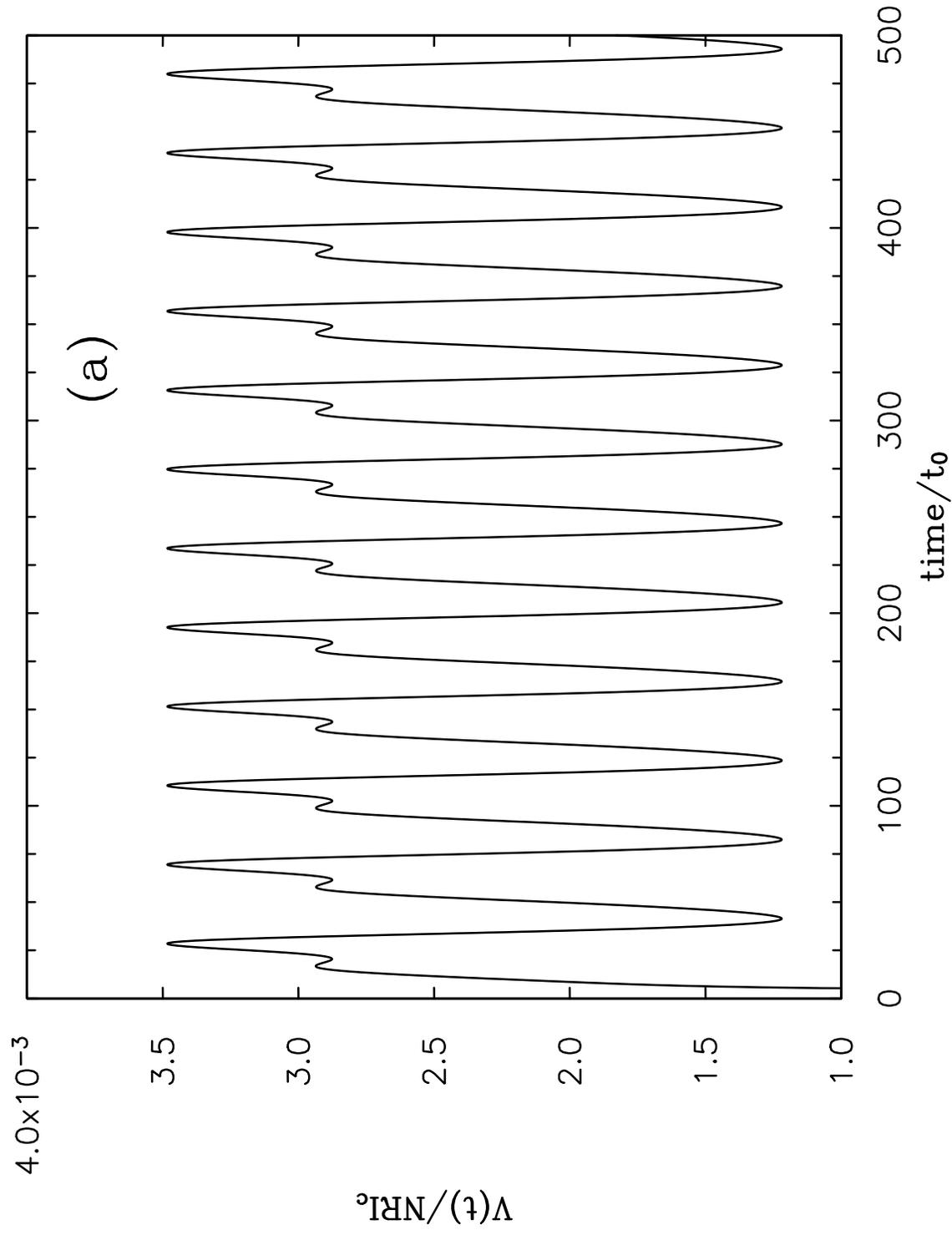
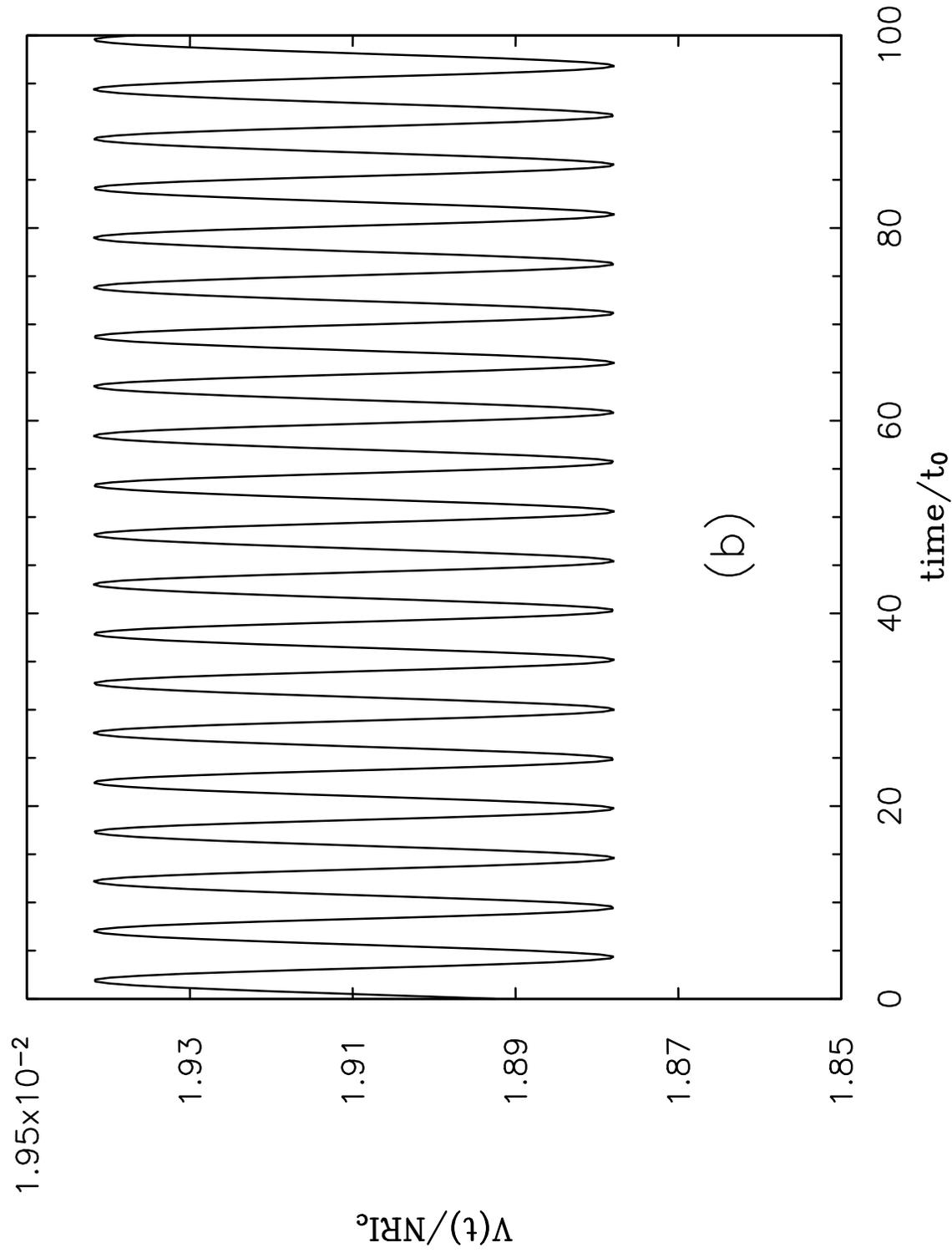
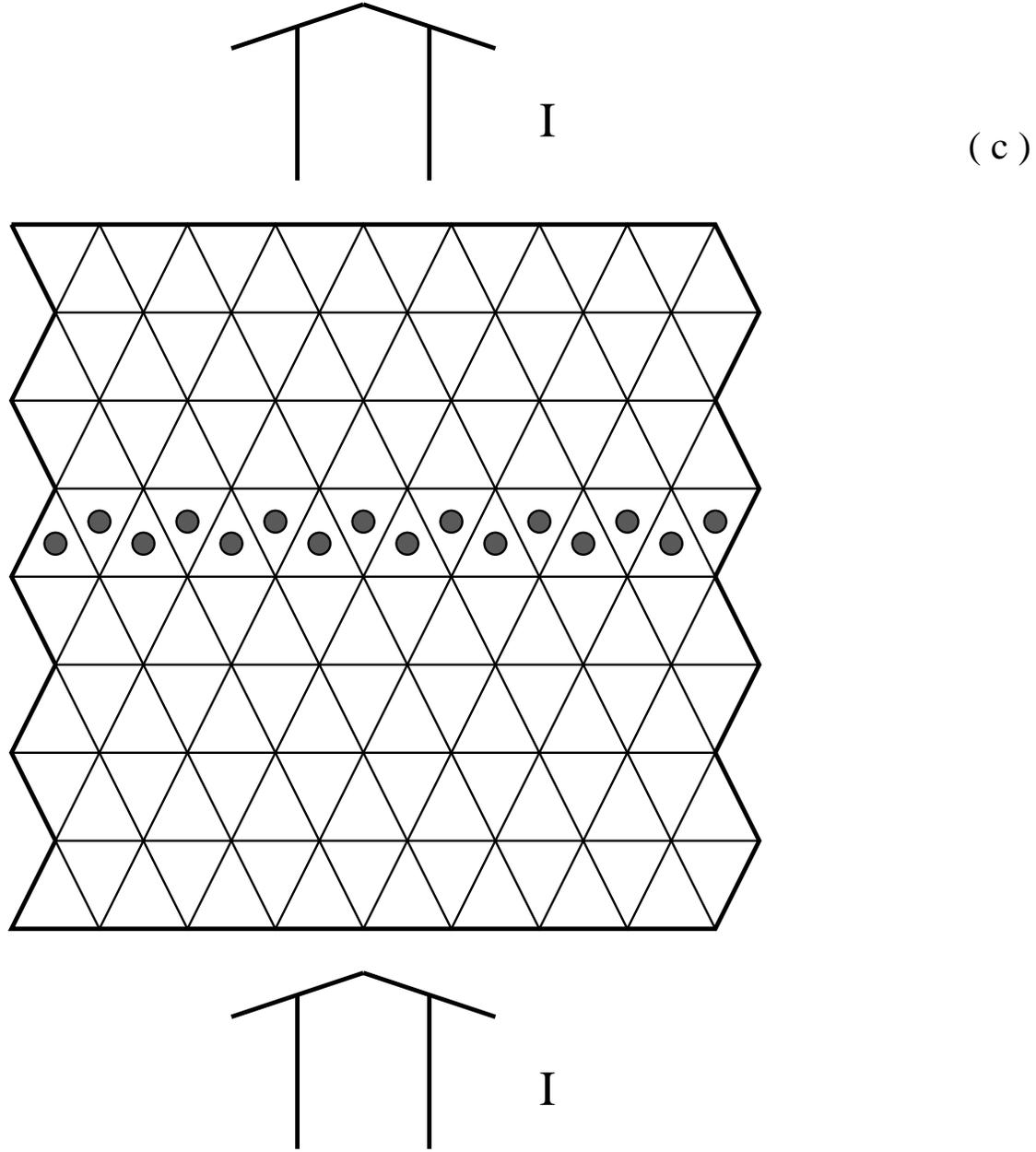
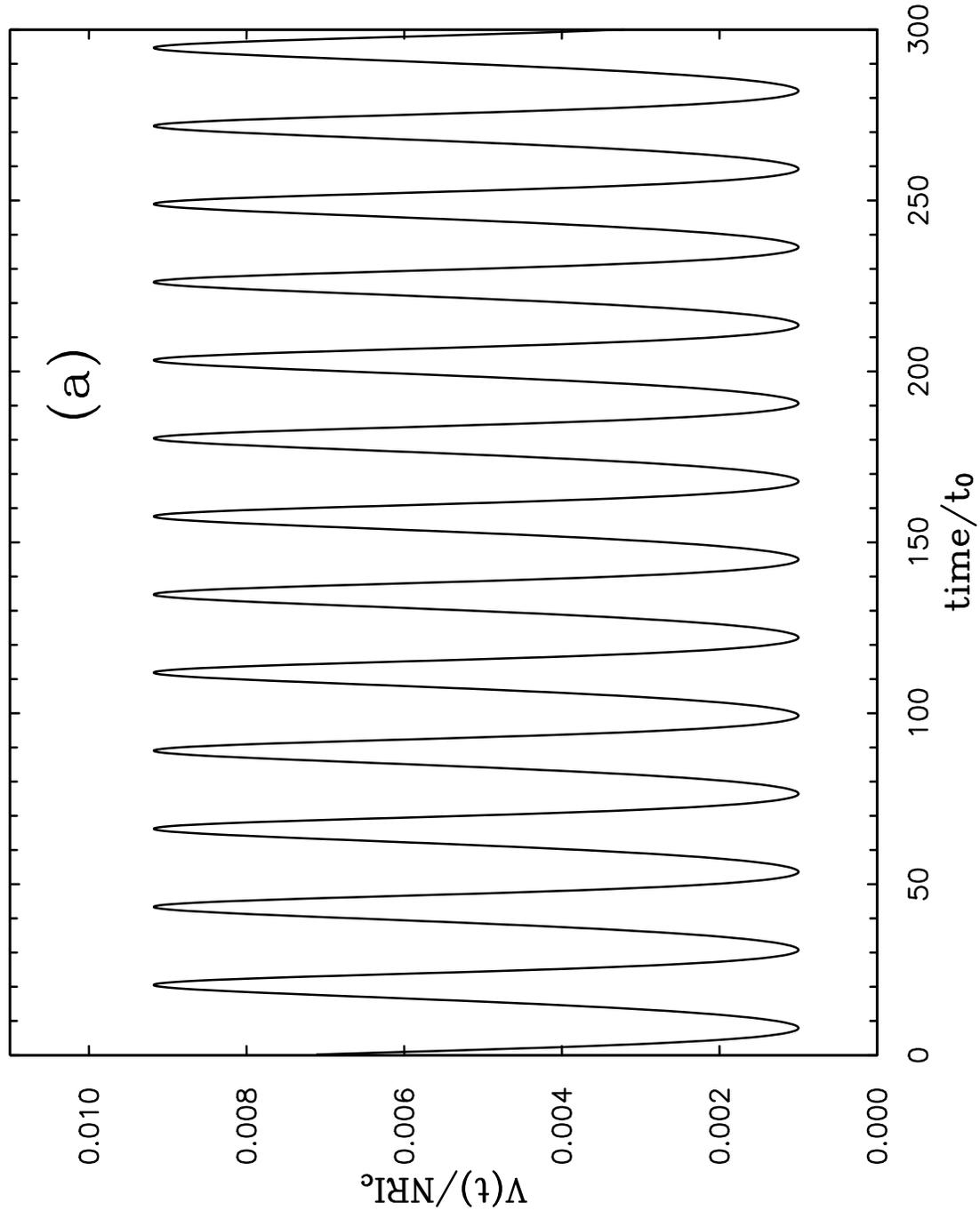
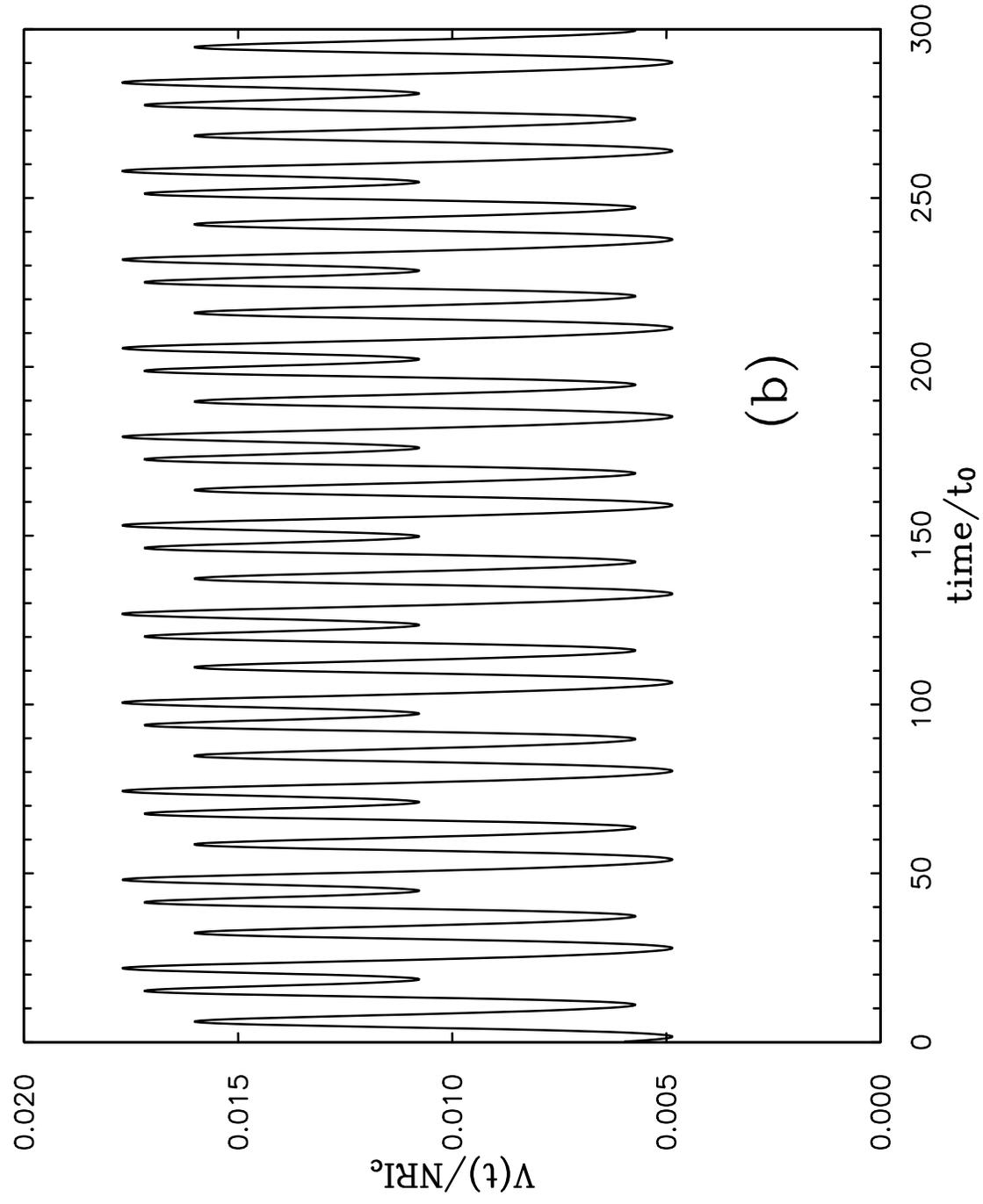
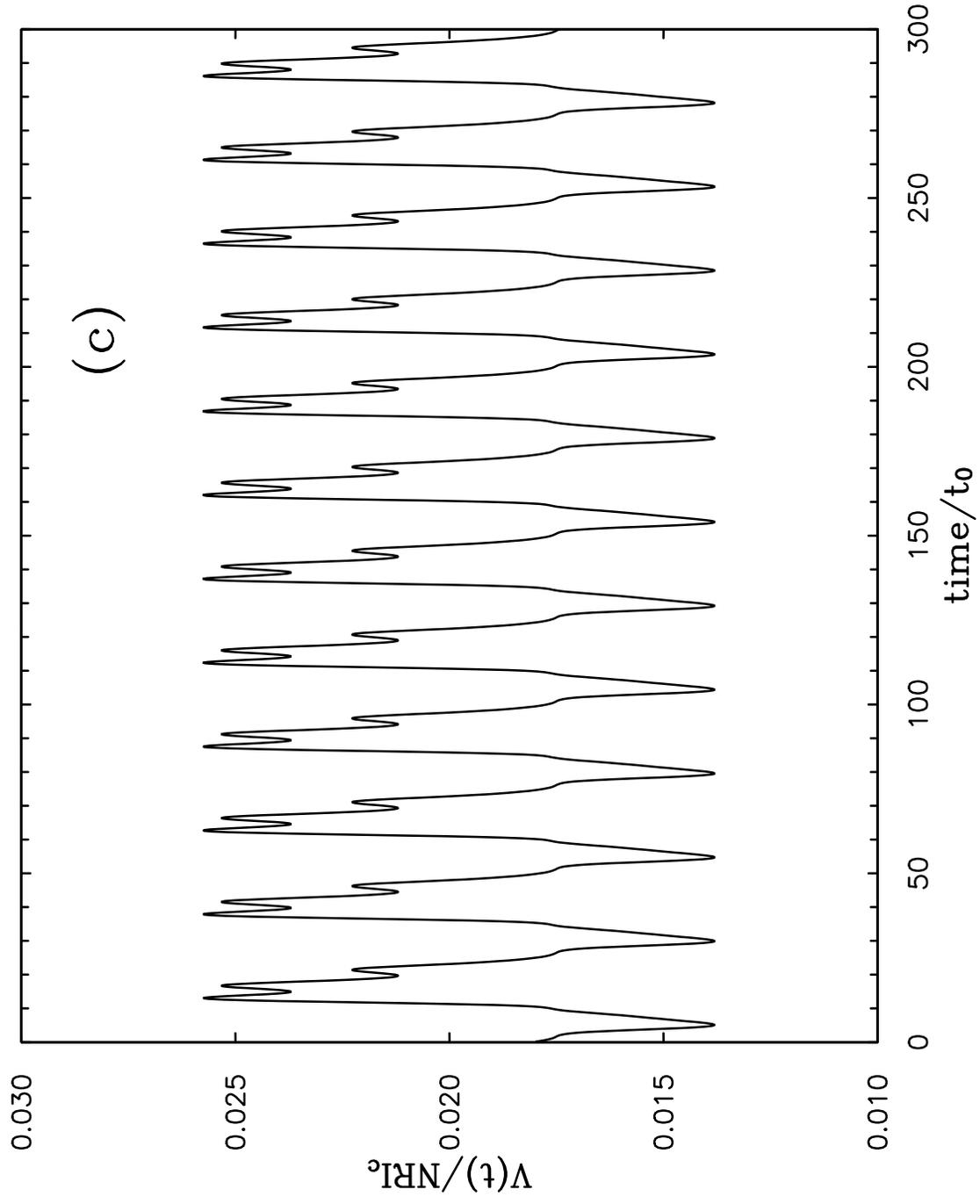
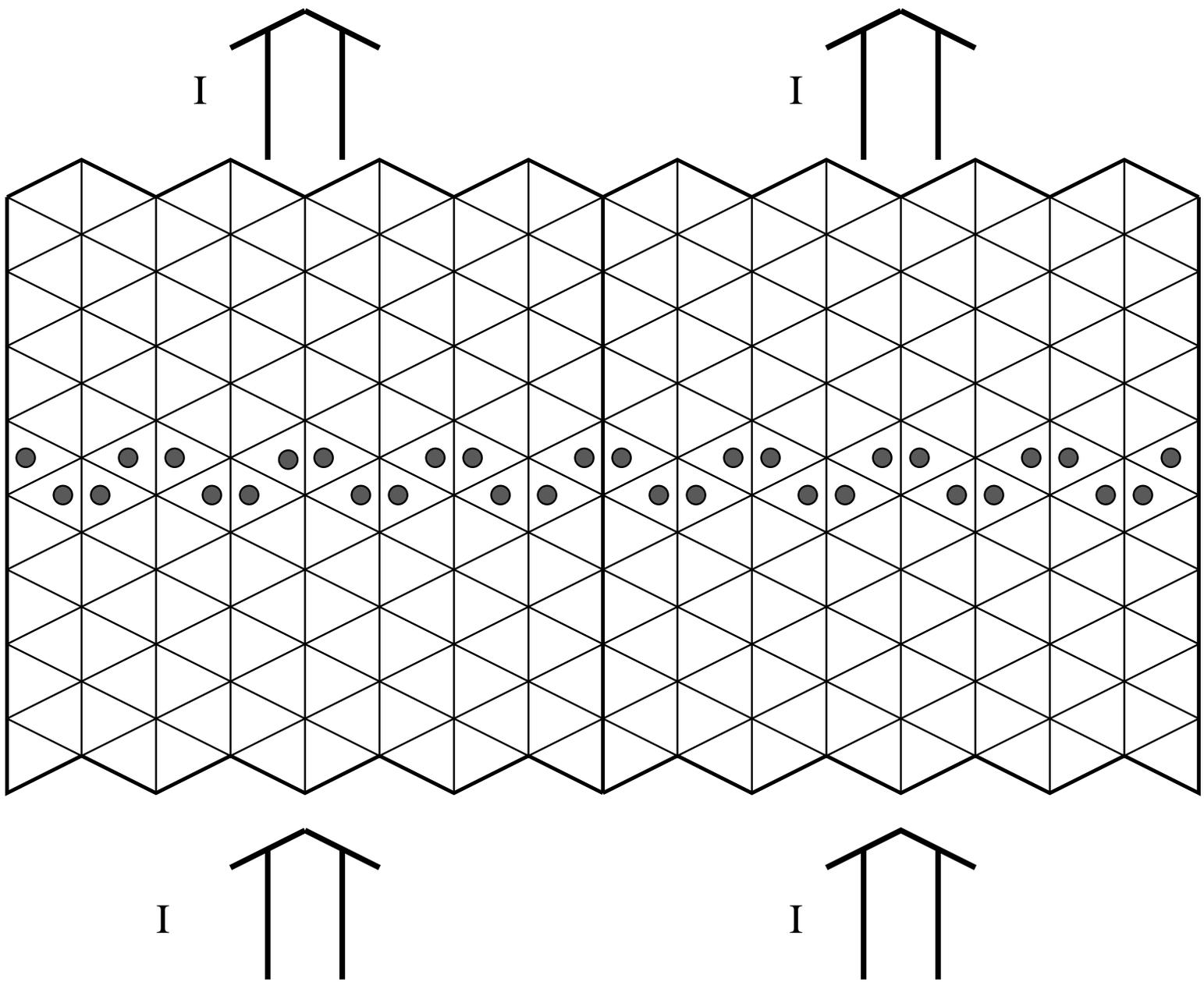
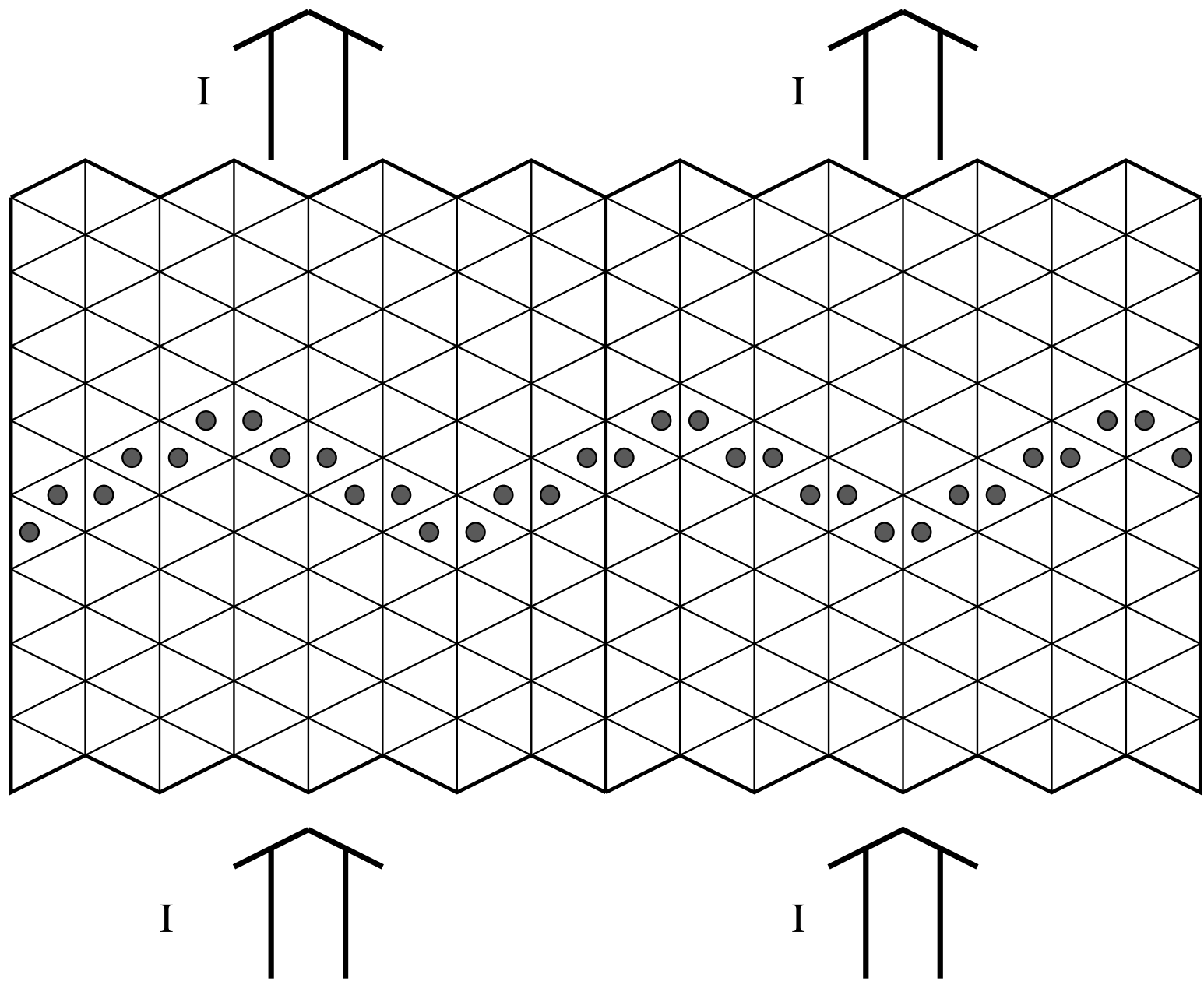
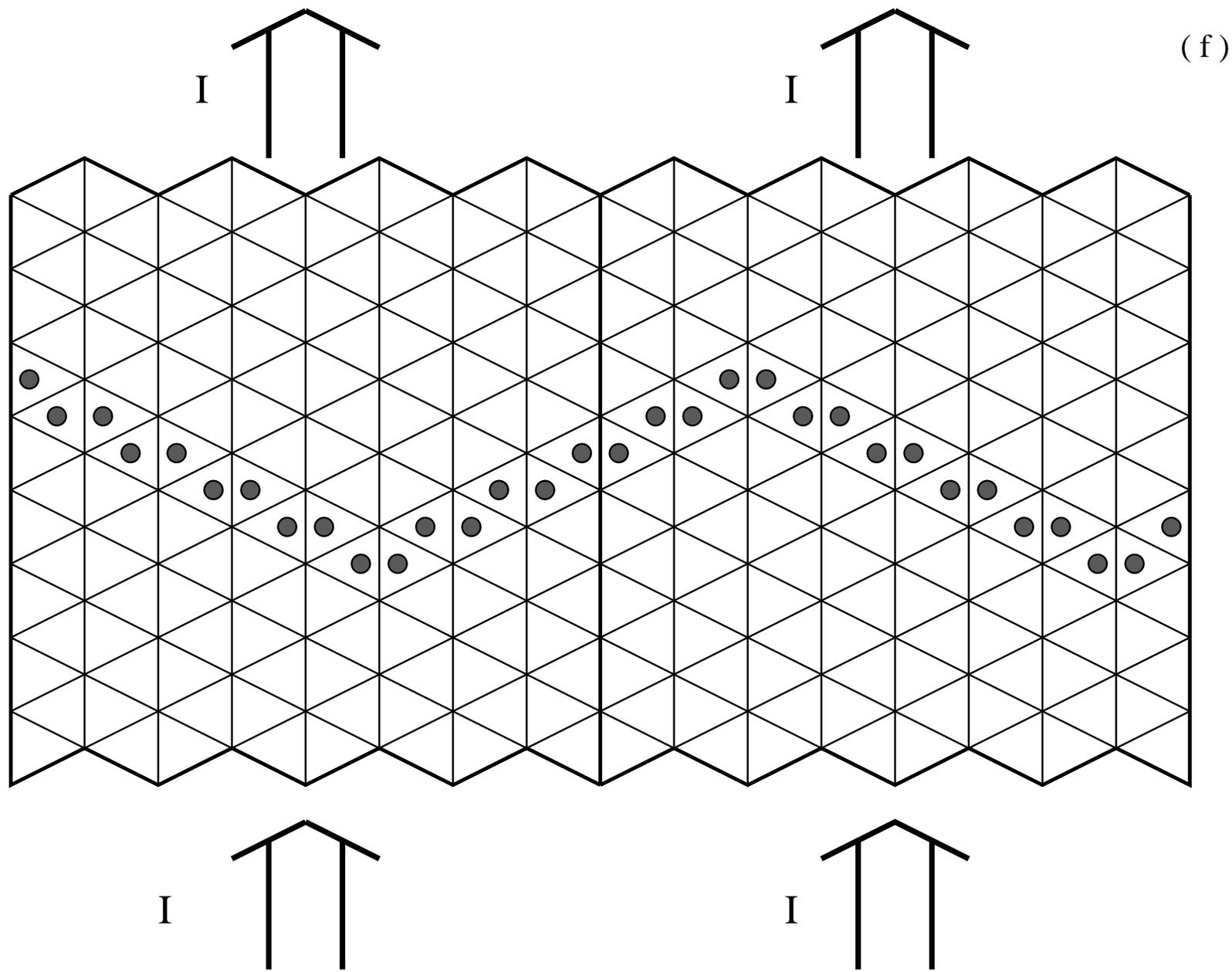
\end{document}